\documentclass{iaus}
\usepackage{graphicx}

\title[IAU 270~~Theory of Cluster Formation: Effects of Magnetic Fields] 
{Theory of Cluster Formation: \\ Effects of Magnetic Fields}

\author[F. Nakamura, \& Z.-Y. Li]  
{Fumitaka Nakamura$^{1,2}$ \and
Zhi-Yun Li$^3$}

\affiliation{$^1$Division of Theoretical Astrophysics,  
National Astronomical Observatory of Japan 
\\ $^2$Institute of Space and Astronautical Science, 
Japan Aerospace Exploration Agency, 3-1-1 Yoshinodai, Sagamihara, 
Kanagawa 229-8510, Japan \\ 
email: {\tt fumitaka.nakamura@nao.ac.jp} 
\\[\affilskip] $^3$Astronomy Department, University of Virginia, 
P. O. Box 400325, Charlottesville, VA 22904 \\
email: {\tt zl4h@virginia.edu}}

\pubyear{2010}
\volume{270}  
\pagerange{1--8}
\jname{Computational Star Formation}
\editors{J. Alves, B. G. Elmegreen, J. M. Girart, \& V. Trimble, eds.}
\begin{document}

\maketitle

\begin{abstract}
Stars form predominantly in clusters inside dense clumps of 
molecular clouds that are both turbulent and magnetized. 
The typical size and mass of the cluster-forming clumps are 
$\sim 1$ pc and $\sim 10^2 - $ 10$^3$ M$_\odot$, respectively.
Here, we discuss some recent progress on numerical simulations 
of clustered star formation in such parsec-scale dense clumps 
with emphasis on the role of magnetic fields.  
The simulations have shown that magnetic fields 
tend to slow down global gravitational collapse and thus star 
formation, especially in the presence of protostellar outflow 
feedback.  Even a relatively weak magnetic field can retard star formation 
significantly, because the field is amplified by supersonic 
turbulence to an equipartition strength. However, in such a 
case, the distorted field component dominates the uniform one.
In contrast, if the field is moderately-strong, the uniform 
component remains dominant. Such a difference in the magnetic 
structure is observed in simulated polarization maps of  
dust thermal emission. Recent polarization measurements show
that the field lines in nearby cluster-forming clumps are 
spatially well-ordered, indicative of a rather strong field. 
In such strongly-magnetized clumps, star formation should 
proceed relatively slowly; it continues for 
at least several global free-fall times of the parent dense 
clump ($t_{\rm ff}\sim $ a few $\times 10^5$ yr). 
\end{abstract}

\firstsection 

\section{Introduction}

It is now widely accepted that most stars form in clusters
(\cite[Lada \& Lada 2003]{lada03}).  Observations have revealed 
that clustered star formation occurs in dense compact clumps of 
molecular clouds that are highly turbulent and magnetized.
In addition, almost all massive stars, which greatly 
impact the interstellar environments and thus galaxy evolution, 
are believed to be produced in clusters.
Thus, understanding how star clusters form is one of the 
central issues in star formation studies.
Although our current knowledge of star cluster formation 
still remains limited, great efforts have been paid 
to clarify the relative importance between magnetic fields 
and turbulence in clustered star formation, 
on the basis of numerical simulations. 
In the present paper, we discuss some recent progress 
on such numerical simulations  with emphasis on the role of magnetic fields.

The dynamical stability of a magnetized cloud is determined by 
the ratio of the mass of a cloud to its magnetic flux.
When the mass-to-flux ratio is larger than the critical value  
$1/(2\pi G^{1/2})$, the magnetic field alone cannot 
support the whole cloud against gravitational collapse. Such a 
{\it magnetically-supercritical} cloud collapses dynamically.
On the other hand, when the mass-to-flux ratio is smaller than the
critical value, the magnetic field can support the whole cloud.
Such a {\it magnetically-subcritical} cloud cannot collapse
without losing magnetic flux.  Since molecular gas is almost 
neutral, ambipolar diffusion can play a role in reducing the 
magnetic flux from dense cores and clumps, leading to 
the global gravitational collapse.  
Thus, it is of great importance to measure the mass-to-flux ratios 
of molecular clouds and their substructures such as clumps and cores, 
in order to assess their dynamical stability, although it is  
difficult to do.

The only technique for directly measuring magnetic field strength
in molecular clouds is through the Zeeman effect of molecular 
lines, which yields the line-of-sight, rather than the total, 
field strength. Cloud mass determination is also uncertain,  
due to uncertainties in the abundances of the observed molecules 
and distances to the clouds. These difficulties make it difficult 
to determine the mass-to-flux ratio accurately. Available 
Zeeman measurements indicate that the median mass-to-flux 
ratios for dense cores of dark clouds 
(\cite[Troland \& Crutcher 2008]{troland08})
and massive star-forming dense clumps 
(\cite[Falgarone et al. 2008]{falgarone08}) are 
within a factor of a few of the critical value (after geometric
corrections). 
Too weak a magnetic field may contradict the observed polarization 
maps of star forming regions, which often indicate
spatially well-ordered magnetic field lines.
In Section \ref{sec:polarization}, 
we present the polarization maps of the dust thermal 
emission from our simulation data.  
Our polarization maps indicate that in the presence of 
a weak magnetic field, the polarization vectors have 
large fluctuations and weak polarization degrees, 
whereas in the presence of a strong magnetic field, 
the vectors are spatially well-ordered.

\vspace*{-2mm}
\section{Setup of Cluster Formation Simulations}

Recent observations have revealed that active cluster-forming 
regions are not distributed uniformly in parent molecular clouds, 
but localized and embedded in dense clumps.
A good example is the nearby well-studied star-forming region, the Perseus
molecular cloud. This cloud has two active cluster-forming regions:
IC 348 and NGC 1333, which contain about 80 \% of the young stars 
associated with this cloud (\cite[Carpenter 2000]{carpenter00}).  
The mass fraction of molecular gas occupied by these two regions is 
only less than a few tens \%.
The typical size and mass of such dense clumps is about 1 pc 
and 10$^2-10^3 M_\odot$, respectively 
(\cite[e.g., Ridge et al. 2003]{ridge03}).
Here, we choose such a parsec-scale dense clump as the initial condition
of our simulations.

The initial cloud is a centrally-condensed
spherical clump with an initial density profile of 
$\rho(r) = \rho_c/[1+(r/r_c)^2]$, where $r_c = L/6$ is the radius of the
central plateau region and $L=2$ pc is the length of the simulation box.
We adopt a central density of $5.0\times 10^{-20}$ g cm$^{-3}$,
corresponding to the central free-fall time $t_{\rm ff,c}=0.30$ Myr.
It yields a total clump mass of $M_{\rm tot}=884 M_\odot$.
The average clump density is $\bar{\rho}=7.5\times 10^{-21}$ g
cm$^{-3}$, corresponding to the global free-fall time 
$t_{\rm ff}=0.77$ Myr.
We assume the isothermal equation of state with a sound speed 
of $c_s=0.23$ km s$^{-1}$ for a mean molecular weight of $\mu=2.33$
and the gas temperature of $T=15$ K.
The periodic boundary condition is applied to each side of a cubic
simulation box.

At the beginning of the simulation, we impose on the cloud a uniform 
magnetic field along the $x$-axis. The field strength is specified by
the plasma $\beta$, the ratio of thermal pressure to magnetic pressure
at the clump center, through $B_0=25.8\beta^{-1/2} \mu$G.  In units of 
the critical value, the mass-to-flux ratio in the central flux tube is
given by $\Gamma_0 = 8.3 \beta^{1/2}$.  The mass-to-flux ratio
for the initial clump as a whole is $\Gamma=3.0\beta^{1/2}$.
In the present paper, we concentrate on the simulation results of three 
models with different magnetic field strength, to discuss the role of
magnetic fields in cluster formation: 
(1) $\Gamma = 3.0\times 10^3$ ($\beta=10^6$), 
(2) $\Gamma = 4.3$ ($\beta=2$), and 
(3) $\Gamma = 1.4$ ($\beta=0.2$).
Following the standard  procedure, we stir
the initial clump at the beginning of the simulation with
a turbulent velocity field of power spectrum $v_k^2 \propto k^{-4}$ and 
rms Mach number ${\cal M}=5$.
Our simulation has a relatively modest resolution of $256^3$. When the
density in a cell crosses the threshold $\rho_{\rm th}=400 \rho_0$,
we create a Lagrangian particle at the position of the maximum density.
The particle is assumed to move with the mass-weighted mean velocity 
calculated from the extracted gas.
The particle is also allowed to accrete the surrounding gas. 
To mimic the effect of protostellar outflow feedback, the particle injects 
into the ambient gas a momentum that is proportional to the particle
mass $M_*$.  The outflow momentum is scaled with the dimensionless 
outflow parameter $f$ as 
$P=f (M_{*}/M_\odot)(V_w/100{\rm km \ s^{-1}})$, 
where we adopt $V_w=100$ km s$^{-1}$ and $f=0.5$.
Each outflow has bipolar and spherical components.
The ratio of the bipolar to spherical components is 0.75.
The direction of the bipolar component is set to be parallel to the local
magnetic field direction.

\begin{figure}[t]
\begin{center}
 \includegraphics[width=\textwidth]{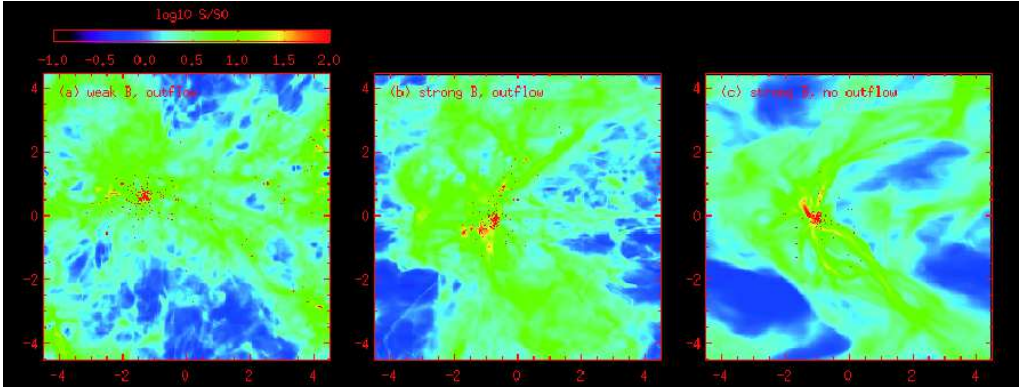} 
 \caption{Snapshots of the column density distribution
for (a) the weakly-magnetized model with the outflow feedback
[$\Gamma=4.3$ ($\beta=2.0$) and $t=2.8 t_{\rm ff}$],
(b) strongly-magnetized model with the outflow feedback, 
[$\Gamma=1.4$ ($\beta=0.2$) and $t=4.0 t_{\rm ff}$]
and (c) strongly-magnetized model without the outflow feedback
[$\Gamma=1.4$ ($\beta=0.2$) and $t=1.3 t_{\rm ff}$], 
at the stage when the star formation efficiency 
has reached 15 \%. 
The initial magnetic field direction is parallel to the horizontal axis.
The small red dots indicate the positions
of stars. The units of length is the Jeans length of 
the initial cloud $L_J = 0.22$ pc and the global free-fall time is
 $t_{\rm ff}=0.77$ Myr.
}
\label{fig:snapshot}
\end{center}
\end{figure}

\section{Numerical Results}

\subsection{Effects of Magnetic Fields}

In Figs. \ref{fig:snapshot}a and \ref{fig:snapshot}b, 
we compare the snapshots of the weakly-magnetized and 
strongly-magnetized models at the stage when the star formation 
efficiency (hereafter SFE) has reached 15 \%.
  For the models in panels (a) and (b),
the outflow feedback is taken into account.
For comparison, we present the snapshot of the strongly-magnetized 
model without the outflow feedback
at the same stage of SFE $=15$ \%.
For all the panels, the initial magnetic field direction is 
parallel to the horizontal axis.
For the weakly-magnetized model, the overall column density
distribution appears insensitive to the initial magnetic 
field direction, implying that the cloud dynamics is 
controlled by supersonic turbulence.  
On the other hand, in the presence of the strong magnetic field, 
the overall density distribution tends to be elongated 
perpendicular to the initial magnetic field direction. 
In the presence of outflow feedback, the column density distribution 
shows many cavities that are created in part by the protostellar outflows.
On the other hand, in the absence of the outflow feedback, 
the cavities are less prominent. This is in part because 
in the absence of the outflow 
feedback, the substantial amount of turbulence has decayed 
and the turbulence is too weak to create many cavities.
Another reason is that the protostellar outflow-driven turbulence 
have more compressible mode that tends to create more cavities
(\cite[see also Carroll et al. 2010]{carroll10}).

\begin{figure}[t]
\begin{center}
 \includegraphics[width= 0.95 \textwidth]{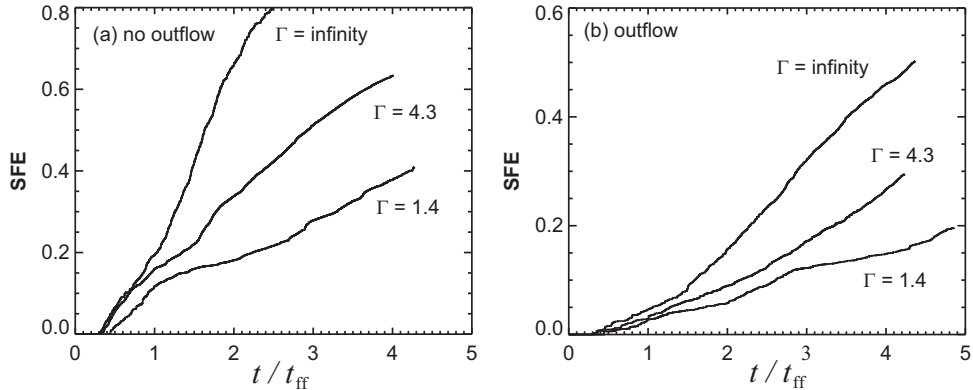} 
 \caption{(a) Star formation efficiencies 
for three models with no outflow feedback
against evolution time that is normalized to the 
global clump free-fall time, 
(1) $\Gamma=3.0\times 10^3$,
(2) $\Gamma=4.3$ (weak magnetic field), and 
(3) $\Gamma=1.4$ (moderately-strong magnetic field). 
Protostellar outflow feedback is not taken into account 
for these models.  
(b) Same as panel (a) but for models with outflow feedback.
}
\label{fig:sfe}
\end{center}
\end{figure}

To illustrate how the initial magnetic field influences 
star formation in the clump, we present in Fig. \ref{fig:sfe}a
the star formation efficiencies against 
the evolution time for the three models with different magnetic field 
strengths, turning the outflow feedback off.
Clearly, the magnetic field tends to retard star formation.
In the period at which the star formation efficiency has reached
15 \% from the formation of the first star, the star formation 
rate per one global free-fall time is estimated to be 29 \%, 
27 \%, and 21 \%, for $\Gamma=3.0\times 10^3$, 4.3, and 1.4, respectively.
The reduction of the star formation rate due to the magnetic field 
is more significant for later stages, although it is 
not enough to reproduce the observed level of a few \%.
This is in good agreement with the results of the recent 
MHD SPH simulations by \cite{price08} and \cite{price09} who followed 
the evolution of 50 $M_\odot$ clumps with $\infty \geq \Gamma \geq 3$
until $t \lesssim 1.5 t_{\rm ff}$,where $\Gamma = \infty$ corresponds to a
nonmagnetized model. 
Although their initial clump mass is too small for 
a cluster-forming clump, their star formation rates per one 
free-fall time are estimated to be over 10 \% in the range of 
$\infty \geq \Gamma \geq 3$.
These simulations imply that other factors are needed to significantly retard 
star formation in a pc-scale dense clump.

\subsection{A Significant Role of Protostellar Outflows in Cluster Formation}
\label{subsec:mag}

\begin{figure}[t]
\begin{center}
 \includegraphics[width=\textwidth]{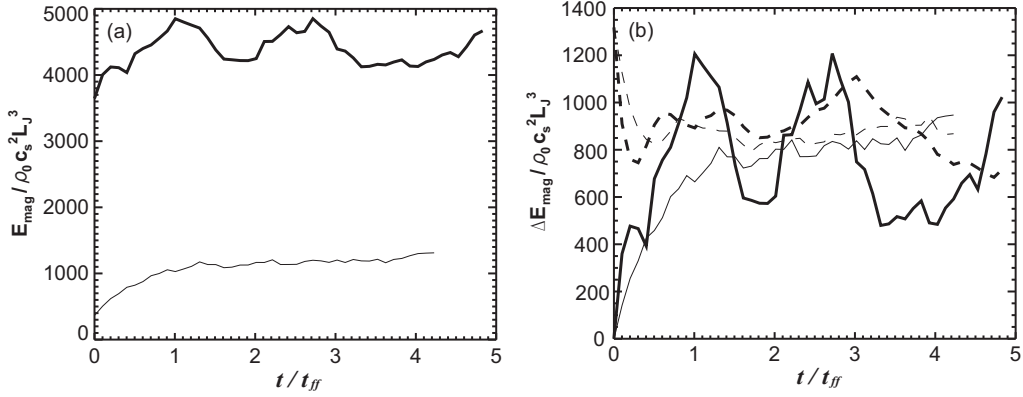} 
 \caption{(a) Time evolution of total magnetic energy for two magnetized
 models. Thick and thin lines are for the models with moderately-strong 
($\Gamma = 1.4$) and weak magnetic fields ($\Gamma = 4.3$), 
respectively.
(b) Time evolution of magnetic energy of amplified component
({\it solid lines}) and kinetic energy of turbulence ({\it dashed
 lines}). The magnetic energy of the amplified component tends to 
approaches the kinetic energy for each model.
}
   \label{fig:mag}
\end{center}
\end{figure}

There are two ways to slow down the star formation rate more
significantly.   One way is the feedback from forming stars.
 In a cluster-forming clump, we usually see stars at
different stages of formation, all in close proximity to each
other, with one generation of stars potentially affecting 
the formation of the next.  
The effects of star formation and its feedback on cluster formation
have been studied by two different groups.
\cite{price09} included radiative feedback from forming stars in their
MHD SPH simulations, and found that the radiative feedback significantly 
suppress the small-scale fragmentation by increasing 
the temperature in the high-density material near the protostars. 
However, it does not much change the global star formation efficiency 
in parent clumps.  
\cite{li06}, \cite{nakamura07}, and \cite{wang10} considered the effects of 
protostellar outflow feedback on cluster formation.  
They found that the protostellar outflow feedback can significantly 
reduce the star formation efficiency, although the moderately-strong 
magnetic fields are necessary to reproduce 
the observed level of low SFEs.
In Fig. \ref{fig:sfe}b, we present the star formation efficiencies
against the evolution time for the same three models as 
in Fig. \ref{fig:sfe}a but including the protostellar outflow feedback.
It is clear that star formation rate is greatly reduced by the inclusion 
of the outflow feedback.  In fact, the star formation rate
per one free-fall time is estimated to be 7.6 \%, 5.6 \%, and 4.6 \%
for the models with $\Gamma=3.0\times 10^3$, 4.3, and 1.4, respectively.
It is interesting to note that even in the presence of the weak magnetic
field, star formation is greatly retarded when the outflow feedback is 
included.  The reason of the great reduction of SFR can be seen in Fig. 
\ref{fig:mag}, where the total magnetic energy is plotted 
against the evolution time.

\begin{figure}[t]
\begin{center}
 \includegraphics[width=\textwidth]{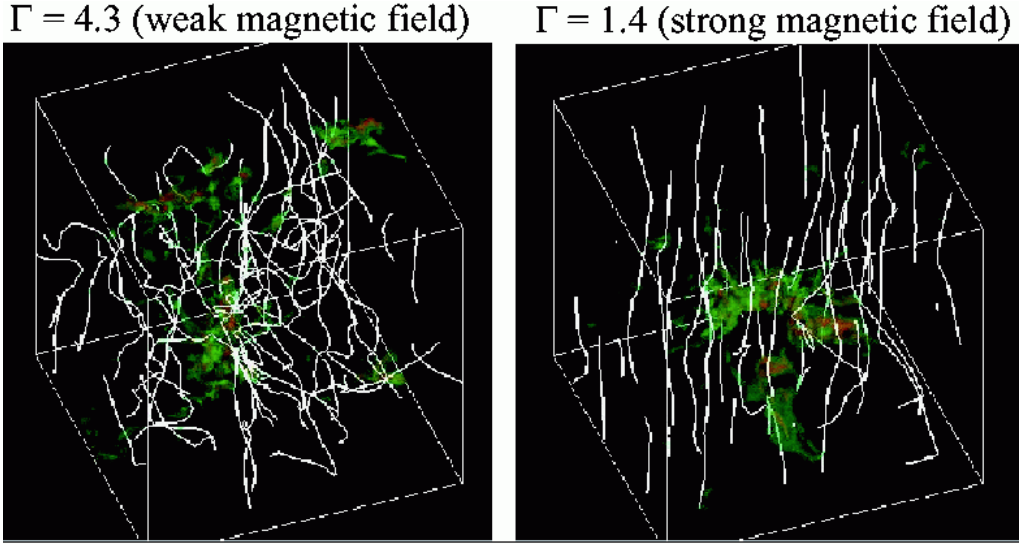} 
 \caption{3D view of the density and magnetic field distributions 
for  the weakly-magnetized model with
$\Gamma=4.3$ ({\it left}) and the strongly-magnetized model with
$\Gamma=1.4$ ({\it right}).
}
\label{fig:3d}
\end{center}
\end{figure}

As shown in Fig. \ref{fig:mag}a, for the strong magnetic field,  
the total magnetic energy is dominated by the background uniform 
field that does not contribute to the force balance at the initial
cloud.  In Fig. \ref{fig:mag}b,  we illustrate  
the time evolution of the magnetic energy stored in the 
distorted component that was amplified by supersonic turbulence. 
Here, we computed the magnetic energy stored in the distorted component
by subtracting the initial magnetic energy from the total magnetic energy.
For both the models, the amplified component increases with time 
and then begins oscillations about a level value, after a free-fall time.
Furthermore, the magnetic energy 
of the amplified component becomes comparable to the kinetic energy 
of the dense gas for each model, indicating that the energies have 
reached an equipartition level. 
For the weaker magnetic field, the amplified component is more 
important than the initial uniform field, resulting in 
a significantly-distorted magnetic field structure 
(left panel of Fig. \ref{fig:3d}).  In contrast, the global field is 
well-ordered for the stronger initial field
(right panel of Fig. \ref{fig:3d}).

\subsection{Cluster Formation in Initially Magnetically-Subcritical Clumps}

\begin{figure}[t]
\begin{center}
 \includegraphics[width=\textwidth]{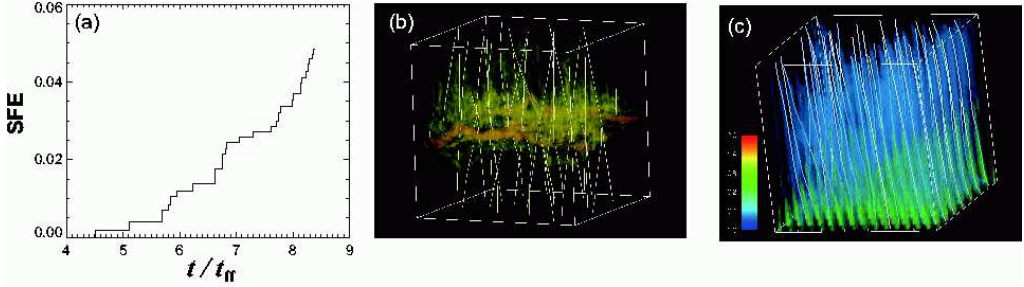} 
 \caption{(a) Star formation efficiency 
for the initially-magnetically-subcritical cloud with 
$\Gamma=0.8$ and the initial turbulent Mach number ${\cal M}=10$ 
against the evolution time that is normalized to the
free-fall time of the gas sheet (see \cite[Nakamura \ Li 2008]{nakamura08}). 
(b) 3D view of density and magnetic field distribution 
of the model presented in panel (a).
The white lines indicate the magnetic field lines.
(c) Blow-up of the low density envelope of the
model presented in panel (b).  Many non-self-gravitating filaments
parallel to the local magnetic field lines are seen.
}
   \label{fig:subcritical}
\end{center}
\end{figure}

Another way to slow down star formation is 
to have a stronger magnetic field. 
\cite{nakamura08} considered a magnetically-subcritical cloud, as 
a model for dispersed, rather than clustered, star formation. 
They performed 3D MHD simulations including ambipolar 
diffusion, which reduces the magnetic flux from the dense regions.  
In Fig. \ref{fig:subcritical}, we present the results of the initially
magnetically-subcritical model with $\Gamma = 0.8$ and ${\cal M}=10$.
In this model, the star formation rate is as small as 0.5 \% in the 
early phase of star formation, but it increases with time and reaches
about 1 \% by the end of the computation when the SFE has reached 
about 5 \%. In other words, the star formation is accelerated.
An unique characteristic of this model is the diffuse filamentary
structure seen in the low-density envelope. Such a filamentary
structure is an important characteristic of MHD turbulence in the
presence of the strong magnetic field.  Such a feature has indeed been
observed in a nearby low-mass star-forming region, the Taurus molecular cloud.
If the cluster-forming clumps are created out of a magnetically-subcritical 
parent molecular cloud, the low star formation rate can be easily
achieved.
One possibility to form the subcritical clumps is the external
large-scale flows induced by turbulence and/or supernovae.
In fact, infrared dark clouds, which are thought to be
regions in the early stages of cluster formation, 
tend to be very filamentary (\cite[Gutermuth et al. 2008]{Gutermuth08}). 
In the presence of strong magnetic fields, the filamentary clumps
are almost perpendicular to the global magnetic fields, and 
diffuse filaments that are parallel to the magnetic field lines 
are likely to be observed in the low-density envelope.
Very recently, Sugitani et al. (2010, in preparation) performed
the polarization observations toward Serpens South, the nearby infrared
dark cloud, and found that the filament is almost perpendicular to the 
global magnetic field, implying that the cloud dynamics 
is controlled by the strong magnetic field.
To clarify the possibility of the formation of cluster-forming clumps
from the subcritical media, future observations of magnetic fields
associated with cluster-forming regions will be needed.

\subsection{Polarization Maps}
\label{sec:polarization}

\begin{figure}[t]
\begin{center}
 \includegraphics[width= 0.95 \textwidth]{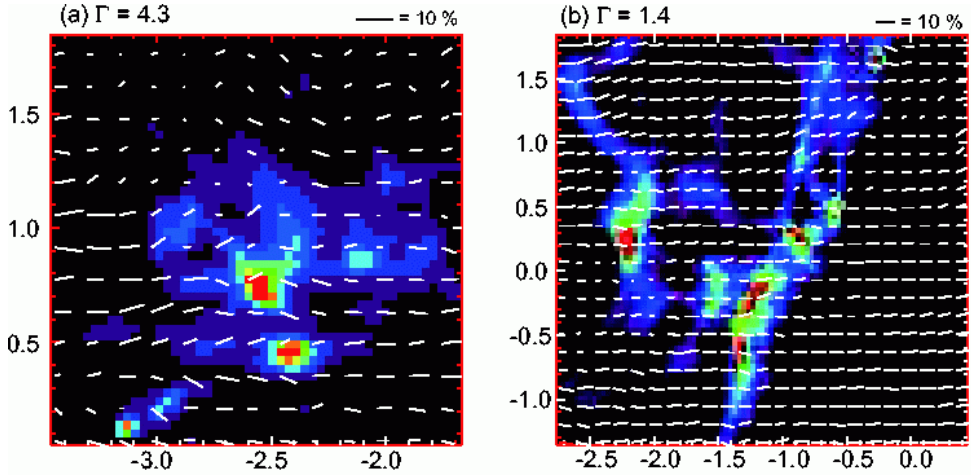} 
 \caption{(a) Polarization maps of the models with 
$\Gamma=4.3$ (weak magnetic field) and 
(b) $\Gamma=1.4$ (moderately-strong magnetic field). 
The length of the polarization vectors is proportional to the degree
of polarization, with the longest vector corresponding to $P=10\%$.
Only one polarization vector is plotted for every four computational 
cells.
The color contour shows the column density distribution. The initial
magnetic field lines are parallel to the horizontal line.
}
\label{fig:polarization}
\end{center}
\end{figure}

Polarization maps of submillimeter thermal dust emission have recently 
been obtained for nearby star forming regions.
Here, we present the polarization maps derived from the simulation data 
for the two magnetically-supercritical models 
with different initial magnetic field strengths
(the same models as described in Section \label{subsec:mag}).
We computed the polarized thermal dust emission from the MHD model
following \cite{padoan01}. We neglect the effect of self-absorption and 
scattering because we are interested in the thermal dust emission at
submillimeter wavelengths.  We further assume that the grain properties
are constant and the temperature is uniform.
The polarization degree is set such that the maximum is equal to 10 \%.
Figure \ref{fig:polarization} shows the dust polarization maps
calculated from the two magnetized models with $\Gamma=4.3$ and 1.4.
Only a small portion of the computation box is shown in each panel
of Fig. \ref{fig:polarization}.
As expected from Fig. \ref{fig:3d}, in the presence of the weak 
magnetic field, the spatial distribution of the polarization vectors 
has relatively large fluctuations.  The polarization degree tends to
be smaller in less dense parts where the magnetic fields are 
strongly distorted.
The column density distribution doesn't show clear filamentary structure.
In contrast, in the presence of the strong magnetic field, 
the filamentary structure is prominent and the filament axes tend to 
be perpendicular to the polarization vectors that are almost parallel 
to the initial magnetic field direction.
Our polarization maps indicate that the polarization observations 
can reflect the magnetic field strengths of the
cluster-forming clumps.

\section{Summary}

We have discussed some recent progress on numerical simulations of
cluster formation with emphasis on the role of magnetic fields.
The numerical simulations indicate that the magnetic field tends to 
slow down star formation significantly. However, it seems 
difficult to slow down star formation at the observed level of a few \%
by the magnetic field alone if the initial field is supercritical
($\Gamma \gtrsim$ a few).
We considered two possibilities to retard star formation to the
observed level.  The first possibility is that the cluster-forming
clumps are created out of the parent magnetically-subcritical cloud.
If the initial field is magnetically subcritical, the
cloud support by the strong magnetic field leads to the great
retardation of the global gravitational collapse. The resultant 
star formation rate decreases significantly.
Another possibility is to slow down the star formation by 
the energy injection due to the protostellar outflows.  
In the presence of the outflow feedback, the star formation is 
greatly reduced even for the 
relatively-weak magnetic fields.  This is because the supersonic 
turbulence amplifies the distorted component of the magnetic fields
significantly for the weak magnetic field.
However, in this case, the magnetic field structure appears to be
random because of the dominant distorted component.  
According to recent polarization observations of cluster-forming regions,
the global magnetic field lines are more or less spatially well-ordered.
For example, \cite{sugitani10} found that the Serpens cloud core 
is penetrated by a hour-glass shaped well-ordered magnetic field and 
is elongated in the cross-field direction.
Very recently, Sugitani et al. (2010, in preparation) also found that
the Serpens South filamentary cloud discovered by \cite{gutermuth08} 
appears to be penetrated by more or less straight global magnetic field.
These observations imply that the magnetic fields associated with the nearby 
cluster-forming regions are likely to be moderately strong.

Moderately-strong magnetic fields are needed to slow down
the internal motions of dense cores where stars form.
Observations of nearby cluster-forming clumps have suggested that 
the internal motions of dense cores tend to be subsonic or
at least transonic 
[e.g., Maruta et al. (2010) for $\rho$ Oph, Saito et al. (2008)].
In our numerical models, the internal motions of the dense cores 
 tend to be subsonic for 
the magnetized models, whereas they are extremely supersonic
for the non-magnetized models (Nakamura \& Li 2010, in preparation).
The moderately-strong magnetic fields (and outflow feedback)
also tend to lower the characteristic mass of the stellar IMF
(Li et al. 2010).

The numerical calculations were carried out mainly on
NEC SX8 at YITP in Kyoto University, and on NEC SX9 at CfCA 
in National Astronomical Observatory of Japan.


\end{document}